# Direct observation of ultrafast exciton formation in monolayer WSe$_2$

*P. Steinleitner, P. Merkl, P. Nagler, J. Mornhinweg, C. Schüller, T. Korn, A. Chernikov\*, and R. Huber\**

*Department of Physics, University of Regensburg, 93040 Regensburg, Germany*

*\*Authors to whom correspondence should be addressed*

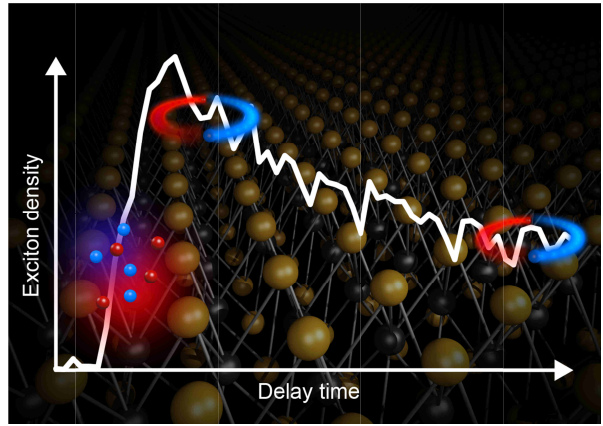

**KEYWORDS: dichalcogenides, atomically thin 2D crystals, exciton formation, ultrafast dynamics.**

**Many of the fundamental optical and electronic properties of atomically thin transition metal dichalcogenides are dominated by strong Coulomb interactions between electrons and holes, forming tightly bound atom-like excitons. Here, we directly trace the ultrafast formation of excitons by monitoring the absolute densities of bound and unbound electron-hole pairs in monolayers of $WSe_2$ following femtosecond non-resonant optical excitation. To this end, phase-locked mid-infrared probe pulses and field-sensitive electro-optic sampling are used to map out the full complex-valued optical conductivity of the non-equilibrium system and to discern the hallmark low-energy responses of bound and unbound pairs. While free charge carriers strongly influence the infrared response immediately after above-bandgap injection, up to 60% of the electron-hole pairs are bound as excitons already on a sub-picosecond timescale, evidencing extremely fast and efficient exciton formation. During the subsequent recombination phase, we still find a large density of free carriers in addition to excitons, indicating a non-equilibrium state of the photoexcited electron-hole system.**

Atomically thin transition metal dichalcogenides (TMDCs) have attracted tremendous attention due to their direct bandgaps in the visible spectral range [1,2], strong interband optical absorption [3,4], intriguing spin-valley physics [5-7], and applications as optically active devices [8-11]. The physics of two-dimensional (2D) TMDCs are governed by strong Coulomb interactions owing to the strict quantum confinement in the out-of-plane direction and the weak dielectric screening of the environment [12,13]. Electrons and holes can form excitons with unusually large binding energies of many 100's of meV [14-19], making these quasiparticles stable even at elevated temperatures and high carrier densities [20,21]. The properties of excitons in 2D TMDCs are a topic of intense research, investigating, e.g., rapid exciton-exciton scattering [22], charged excitons and excitonic molecules [23,24], ultrafast recombination dynamics [19,25-27] or efficient coupling to light and lattice vibrations [4,19,28,29].

In many experiments excitons are created indirectly through non-resonant optical excitation or electronic injection, which may prepare unbound charge carriers with energies far above the exciton resonance [8,18]. Subsequently, the electrons and holes are expected to relax towards their respective band minima and form excitons in the vicinity of the fundamental energy gap. In principle, strong Coulomb attraction in 2D TMDCs should foster rapid exciton formation. Recent optical pump-probe studies have reported characteristic times on the order of 1 ps [30]. The relaxation of large excess energies, however, requires many scattering processes, which lead to non-equilibrium carrier distributions and a mixture of excitons and unbound electron-hole pairs. Hence the question of how excitons and free charge carriers evolve after above-bandgap excitation is of central importance for our fundamental understanding of 2D TMDCs.

Quantifying the densities of bound and unbound carrier populations has remained challenging for optical interband spectroscopy, since both species tend to induce similar modifications in the interband response. Conversely, terahertz and mid-infrared (mid-IR) probes may sensitively discriminate between bound and unbound states via their low-energy elementary excitations [19,31-36]. As schematically illustrated in Figure 1a, excitons efficiently absorb radiation in the spectral range of the intra-excitonic resonances, corresponding to dipole-allowed transitions from the exciton ground state ($1s$) to higher excited states ($2p$, $3p$, $4p$…), labeled in analogy to the hydrogen series. This absorption occurs irrespective of the exciton center-of-mass momentum and interband optical selection rules [19,32-37]. In contrast, free electrons and holes feature a Drude-like response. By using ultrabroadband electro-optic detection of the waveform of a few-cycle mid-IR probe pulse before and after interaction with the electron-hole system, both real and imaginary part of the system conductivity can be independently traced on the femtosecond scale, without resorting to a Kramers-Kronig analysis [31]. Drude and exciton signatures have been clearly identified in bulk $Cu_2O$ [33,37,38] and GaAs quantum wells [32-35]. Recently, this technique has become sensitive enough to probe the intraexcitonic transitions in TMDC monolayers [19,36]. Yet the dynamics of exciton formation in a 2D TMDC has not been studied by direct probing, to the best of our knowledge.

Here, we employ field-sensitive mid-IR femtosecond probing to directly monitor the dynamics of photoexcited electron-hole pairs in the prototypical 2D TMDC WSe$_2$. After highly non-resonant (excess energy > 1.5 eV) femtosecond interband excitation, the complex-valued mid-IR conductivity indicates a rapid carrier relaxation towards the respective band minima during the first 100's of femtoseconds. Remarkably, more than half of the carriers are bound in excitons already 0.4 ps after the excitation. The ratio between excitons and unbound electron-hole pairs increases slightly in the subsequent 0.4 ps and both populations decay on a timescale of a few picoseconds while a significant fraction of free carriers is still observed after 5 ps, characteristic of a highly non-equilibrium electron-hole system.

Our samples were prepared by mechanical exfoliation of WSe$_2$ bulk crystals (HQgraphene) on viscoelastic substrates and were subsequently transferred onto CVD diamond [39]. Monolayer flakes with typical diameters of the order of 100 μm were identified by photoluminescence and reflectance contrast spectroscopy. An overview of the basic experimental concept is schematically illustrated in the inset of Figure 1b. The samples were optically excited with 100 fs laser pulses (repetition rate: 0.4 MHz) centered at a photon energy of either 1.67 eV, for resonant creation of the 1$s$ A exciton of WSe$_2$, or 3.04 eV, for excitation far above the fundamental bandgap (see Supporting Information Figure S1). The pump fluence was set to 19 or 38 μJ/cm$^2$ as detailed below. As a probe, we used a phase-locked mid-IR pulse (duration: 50 fs, FWHM), generated by optical rectification of the fundamental laser output in a 50 μm thick AgGaS$_2$ crystal (see Supporting Information Figure S2). The probe spectrum covered a frequency window between 30 and 53 THz, encompassing the energy of the 1$s$-2$p$ intra-excitonic transition of WSe$_2$ monolayers at ~170 meV [19]. For a given delay time $t_{PP}$ between pumping and probing, the complete electric field waveform $E_{ref}(t_{EOS})$ of the transmitted probe pulse in absence of excitation as well as its pump-induced change $\Delta E(t_{EOS}, t_{PP})$ was recorded electro-optically as a function of the detection time $t_{EOS}$ (see Supporting Information Figure S3). All experiments were performed at room temperature and ambient conditions. We note that the illumination of the sample with a photon energy of 3.04 eV introduced additional broadening and a small redshift of the exciton resonance. The exposure time was thus carefully chosen such that the characteristic intra-excitonic fingerprint was still observed at resonant excitation conditions after illumination with 3.04 eV photons (see Supporting Information Figure S5 and S6 together with discussion).

In the first set of experiments, we tuned the pump photon energy to the interband absorption maximum corresponding to the 1$s$ state of the A exciton (Figure 1b, black arrow) [40]. The corresponding single-particle transition occurs at the K (or K') point of the Brillouin zone [41] (Figure 1c). Figure 2a (upper panel) shows the waveform of the transmitted probe pulse $E_{ref}(t_{EOS})$ in absence of excitation (black curve) together with the pump-induced change $\Delta E(t_{EOS}, t_{PP} = 75$ fs) (red curve). The observed phase shift of $\Delta E$ with respect to $E_{ref}$ of exactly $\pi$ is a hallmark of a dominantly absorptive response, as expected in the case of purely excitonic population [19,42]. This assignment is corroborated by the extracted changes in the real parts of the mid-IR conductivity ($\Delta\sigma_1$) and the dielectric function ($\Delta\varepsilon_1$), presented in Figure 2a (lower panel) roughly corresponding to absorptive and inductive components, respectively. In particular, a broad peak in $\Delta\sigma_1$, centered at a photon energy of $\hbar\omega = 170$ meV, combined with a dispersive shape of $\Delta\varepsilon_1$, crossing the zero-axis at the same energy, are characteristic of the intra-excitonic 1$s$-2$p$ resonance in WSe$_2$, as discussed in detail in ref 19.

A qualitatively different picture is obtained for non-resonant excitation. The pump photon energy of 3.04 eV (Figure 1b, blue arrow) allows for carrier injection into higher-lying states of the WSe$_2$ monolayer, with a variety of possible electronic transitions across the Brillouin zone, as schematically illustrated in Figure 1c. This broad energy region is often chosen for the excitation in optical experiments due to the high absorption of the TMDC materials [40] and the availability of commercial laser sources in this spectral range. The differences to the situation after resonant excitation are already apparent in the electric field trace of the probe pulse, presented in Figure 2b (upper panel). The relative phase of $\Delta E(t_{EOS}, t_{PP} = 400$ fs) deviates strongly from $\pi$ and is closer to $\pi/2$, indicating the predominantly inductive response of an electron-hole plasma [42]. The shape of the corresponding spectral features (Figure 2b, lower panel) is in stark contrast to the observations for resonant injection. Instead of a peak, $\Delta\sigma_1$ is now spectrally flat, accompanied by a negative, monotonically increasing $\Delta\varepsilon_1$ across the experimentally accessible energy range. This shape is characteristic of a Drude-like behavior [31,32,35,42].

For a quantitative analysis of the measured spectra, we apply a phenomenological Drude-Lorentz model [35,42]. Within this two-fluid approach, pump-induced changes in the frequency-dependent dielectric function $\Delta\varepsilon(\omega) = \Delta\varepsilon_1 + i\Delta\sigma_1/(\varepsilon_0\omega)$ are described using two components:

$$\Delta\varepsilon(\omega) = \frac{n_X e^2}{d\varepsilon_0 \mu} \times \frac{f_{1s,2p}}{\frac{E_{res}^2}{\hbar^2}-\omega^2-i\omega\Delta} - \frac{n_{FC} e^2}{d\varepsilon_0 \mu} \times \frac{1}{\omega^2+i\omega\Gamma} \tag{1}$$

The first term, a Lorentzian resonance, accounts for the intra-excitonic 1s-2p absorption. It includes the 1s exciton density $n_X$, the corresponding reduced mass $\mu = m_e m_h / (m_e + m_h)$, obtained from the effective masses $m_e$ and $m_h$ of the constituting electron and hole, the effective thickness $d$ of the monolayer (treated as a thin slab in this model), the oscillator strength $f_{1s,2p}$ of the intra-excitonic transition, the resonance energy $E_{res}$ and the linewidth $\Delta$. Additional constants are the electron charge $e$ and the vacuum permeability $\varepsilon_0$. The second term represents the Drude response of the electron-hole plasma, which depends on the pair density of free carriers $n_{FC}$ and their scattering rate $\Gamma$. For the analysis of the data, we fix $\mu = 0.17\ m_0$ [12] and $f_{1s,2p} = 0.32$ [19], corresponding to electron-hole properties at the K and K' valleys in WSe$_2$, and set the effective layer thickness $d$ to 0.7 nm. The remaining parameters are then extracted by fitting the experimental data. Note that the fact that both independently measured $\Delta\sigma_1$ and $\Delta\varepsilon_1$ spectra need to be simultaneously reproduced poses strict limits to the allowable values of the fitting parameters. The numerical adaptation (Figures 2a and 2b, black dashed curves) yields an overall good fit quality, allowing for a meaningful extraction of the parameters.

For resonant excitation, we find an exciton density of $n_X = 3.4 \times 10^{12}$ cm$^{-2}$ and – as expected – virtually no contribution from unbound electron-hole pairs. The 1s-2p resonance energy of 167 meV and the peak linewidth of 99 meV also compare well with previous observations [19]. Conversely, the response after non-resonant excitation includes a significant fraction of free electron-hole pairs $n_{FC}$ of $0.8 \times 10^{12}$ cm$^{-2}$ (with $n_X = 1.5 \times 10^{12}$ cm$^{-2}$, $E_{res} = 180$ meV, $\Delta = 182$ meV, and $\Gamma = 0.0083$ fs$^{-1}$). Whereas the qualitative shape of the measured response functions in Figure 2b can be roughly accounted for by the Drude model alone (red dotted lines), the two-fluid model (see eq 1) reproduces the data considerably more accurately. This observation already indicates that electron-hole correlations are important even for delay times as short as $t_{PP} < 1$ ps.

The ultrafast evolution of the mid-IR response is systematically shown in Figure 3, which displays spectra of $\Delta\sigma_1$ and $\Delta\varepsilon_1$ for a series of delay times $t_{PP}$ together with the fit curves from the two-fluid model. The corresponding field-resolved time-domain data are given in Supporting Information Figure S7. At $t_{PP} = 0$ fs, we recover a predominantly Drude-like response. After 400 fs, the overall magnitude of the signal increases and a broad resonance centered at about 160 meV develops. Based on the close correspondence to the 1s-2p absorption measured under resonant excitation conditions, this resonance is attributed to the rising exciton population. For $t_{PP} > 0.8$ ps, all pump-induced changes decrease, indicating pair recombination. More importantly, the measured response remains characteristic of a mixture of excitons and free carriers, at all delay times. In particular, we observe a broad excitonic feature in $\Delta\sigma_1$, but also a flat negative response in $\Delta\varepsilon_1$ across all spectra.

The densities of bound and unbound electron-hole pairs, extracted from the Drude-Lorentz fit are presented in Figure 4 (spheres) as a function of $t_{PP}$ together with the total pair density, $n_{tot} = n_X + n_{FC}$. Interestingly, the extracted densities reach their respective maxima at finite delays of $t_{PP} = 0.4$ ps, for $n_{FC}$ and $n_{tot}$, and $t_{PP} = 0.8$ ps for $n_X$. All densities decay on a few-ps scale. For a consistency check, we also measured the transient changes of the electric field $\Delta E(t_{PP})$ at fixed electro-optic delays of $t_{EOS} = 0$ and 6 fs. Since these times correspond to phase-shifts of $\pi$ and $\pi/2$ with respect to the waveform $E_{ref}$ (Figures 2a and 2b), signals $\Delta E$ at these delays should be sensitive to exciton and plasma densities, respectively, as discussed above (see also Supporting Information Figure S4 together with discussion). Indeed, these data (Figure 4, solid curves) almost perfectly trace the dynamics obtained from the full spectral analysis (Figure 4, spheres). The temporal evolution of the remaining fitting parameters (Δ, $E_{res}$, Γ) is given in Supporting Information Figure S8.

Our experimental findings provide important insights into key aspects of the microscopic dynamics of the photogenerated electron-hole pairs. We start out by discussing the delayed rise of the carrier densities with respect to the pump pulse, which we attribute to ultrafast carrier relaxation. As indicated in Figure 1c, the carriers are injected into comparatively flat sections of the electronic bandstructure, defining states with large effective masses. As we model the mid-IR response using the reduced mass from the band edge K (and K') states, the extracted carrier density underestimates the actual number of carriers at these early delay times. This argument is quantitatively consistent with an estimate of the injected carrier density by taking into account an effective optical absorption of the diamond-supported $WSe_2$ monolayer of about 5% at a photon energy of 3.04 eV. As shown in Figure 4, the number of absorbed photons ($4 \times 10^{12}$ cm$^{-2}$) closely matches the maximum measured density $n_{tot} = n_X + n_{FC}$ of bound and unbound electron-hole pairs extracted from the experiment. Thus, at $t_{PP} = 0.4$ ps, we can quantitatively account for every absorbed photon resulting in an electron-hole pair residing close to the band edges of $WSe_2$.

Secondly, for $t_{PP} = 0.4$ ps, roughly 60% of the injected carriers are found to be already bound in excitons. Until $t_{PP} = 0.8$ ps, additional excitons are formed from free charge carriers and the exciton fraction reaches its maximum of about 70% while the plasma density decreases. The exciton formation in monolayer $WSe_2$ is thus significantly faster, by about two orders of magnitude, than the corresponding process in GaAs quantum wells [32,42]. It is also more rapid than the recently reported trion formation in monolayer TMDCs, which evolves on the timescale of several picoseconds [43]. Subsequently, both excitons and unbound carriers decay through radiative and non-radiative recombination. Remarkably, the fraction of bound and unbound carriers remains roughly constant during the carrier lifetime, up to the maximum studied time delay of 5.4 ps. By assuming a thermal population and using the Saha equation [42], the corresponding effective carrier temperature is estimated to be of the order of 1500 K. Consequently, the exciton and plasma populations are far from equilibrium (lattice temperature: 295 K). This scenario appears reasonable considering the high initial excess energy and the short carrier lifetime of a few picoseconds.

Finally, we note, that excitons can be present both as optically bright, spin-allowed K-K excitons (labeled by the location of the corresponding interband transitions in the Brillouin zone) and as spin-forbidden dark states. In addition, relaxation of the exciton population towards inter-valley K-K' [44] and K-Q/Σ states [29] is possible. However, due to the similar binding energies of the different states and the broadening of measured 1$s$-2$p$ resonance, the excitons in our experiment should be generally considered as a mixture of these configurations. Thanks to the relatively small variation of effective exciton masses, ranging from 0.16 $m_0$ (bright K-K excitons) to 0.23 $m_0$ (x-y averaged K-Q excitons) [41], the extracted total density should deviate by less than 26% from the actual pair density in the mixture of different states.

In summary, we have studied the ultrafast intraband response of optically *resonantly* and *non-resonantly* pumped WSe$_2$ monolayers. Field-resolved detection of the spectrally broad mid-IR probe pulse allowed us to monitor the population dynamics of the electron-hole plasma and the tightly bound excitons, individually. Remarkably, we find an exciton fraction as high as 60% after the initial relaxation of the photoexcited carriers towards the band-edge states within the first few 100's of fs. During the subsequent decay of the population on a timescale of several picoseconds, about 70% of the carriers are bound into excitons and the rest of the photoexcited electrons and holes is still present as a free plasma, indicating long-lived non-equilibrium conditions. The findings are of major importance both for our understanding of the fundamental physics of photoexcited 2D TMDCs and for their potential future applications. On the fundamental side, the clear evidence of a rapid exciton formation implies highly efficient exciton injection even under strongly non-resonant excitation conditions. With respect to applications, the presence of a significant fraction of free charge carriers at comparatively long timescales has major implications for the use of TMDC monolayers in future optoelectronic devices, such as sensors, detectors, and photovoltaics.

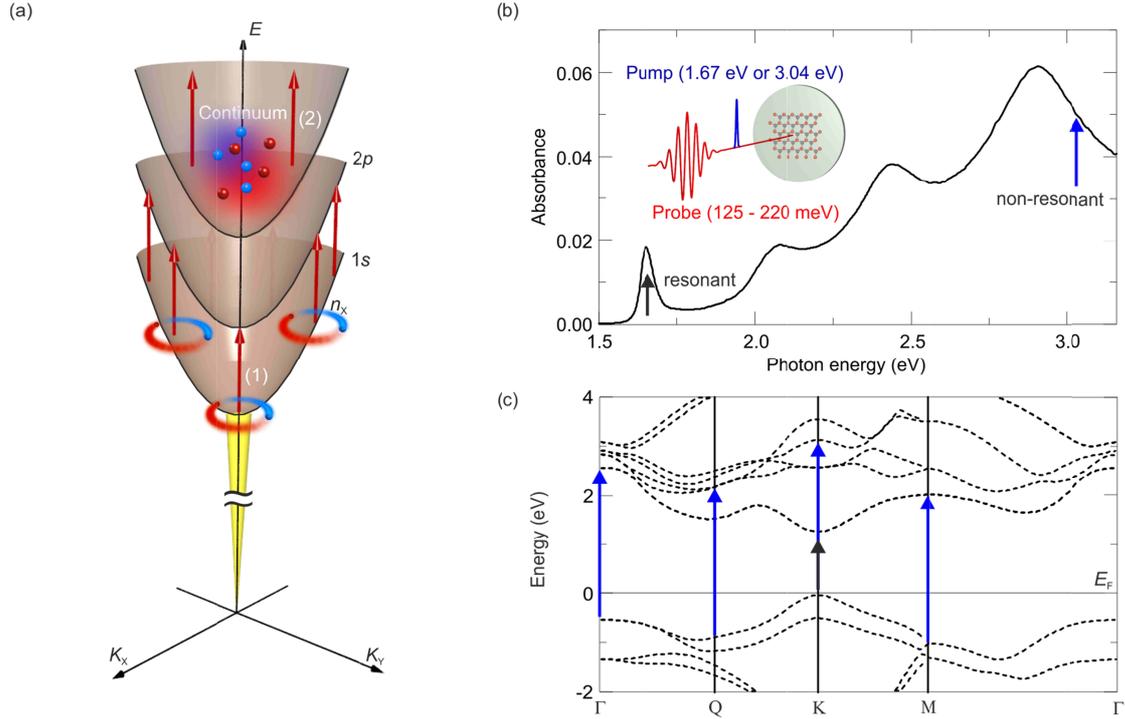

**Figure 1.** (a) Schematic illustration of the low-energy response of excitons and free electron-hole pairs. The dispersion relations are shown in the two-particle picture, corresponding to exciton states with the principal quantum numbers $n$ = 1, 2, 3, etc. and the electron-hole continuum (top-most band), presented as a function of the center-of-mass momentum $K$. The yellow-shaded area represents the region of the light cone, where the excitons can be directly excited by photons due to momentum conservation. Red arrows schematically indicate relevant excitation processes: (1) dipole-allowed transition between 1$s$ and 2$p$ excitons; $n_X$ denotes the 1$s$ population (red and blue spheres with tail); (2) off-resonant excitation of free electron-hole pairs of density $n_{FC}$ (red and blue spheres with halo). (b) Typical optical inter-band absorption spectrum (black solid line) of the WSe$_2$ monolayer [40]. The black arrow marks the photon energy of 1.67 eV for resonant carrier injection, whereas the blue arrow denotes the photon energy used for non-resonant excitation at 3.04 eV. Inset: Schematic of the femtosecond optical-pump / mid-IR-probe experiment of single-layer WSe$_2$ on a diamond substrate. The mid-IR probe pulse (red) is delayed with respect to the optical pump pulse (blue) by a variable time delay. The electric field of the probe is subsequently detected as a function of the electro-optic sampling delay time. (c) Calculated band structure of the WSe$_2$ monolayer from ref 41. The black arrow highlights the fundamental transition for resonant carrier injection. A few possible transitions in case of the non-resonant excitation are sketched by blue arrows. The Fermi energy $E_F$ is set to zero.

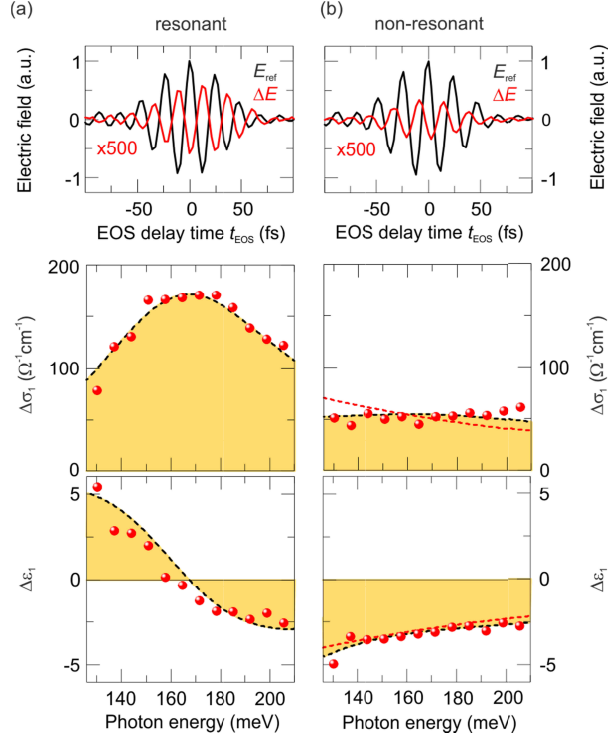

**Figure 2.** (a) Upper panel: time-resolved waveforms of the probe pulse $E_{\text{ref}}$ transmitted through the WSe$_2$ monolayer in the absence of excitation (black curve) and the pump-induced change $\Delta E$ (red curve, scaled up by a factor of 500) after resonant excitation ($t_{\text{PP}}$ = 75 fs, pump fluence: 19 µJ/cm²) in ambient conditions. Lower panel: corresponding pump-induced changes of the real parts of the optical conductivity $\Delta\sigma_1$ (top) and the dielectric function $\Delta\varepsilon_1$ (bottom) as a function of the photon energy. The experimental data is shown by red circles; the results from the Drude-Lorentz model are plotted as black-dashed lines with shaded areas. (b) Same as (a) for non-resonant excitation ($t_{\text{PP}}$ = 400 fs, pump fluence: 19 µJ/cm²). The lower panel includes a fit by the Drude model (i.e., excluding excitons), indicated by the red-dashed line.

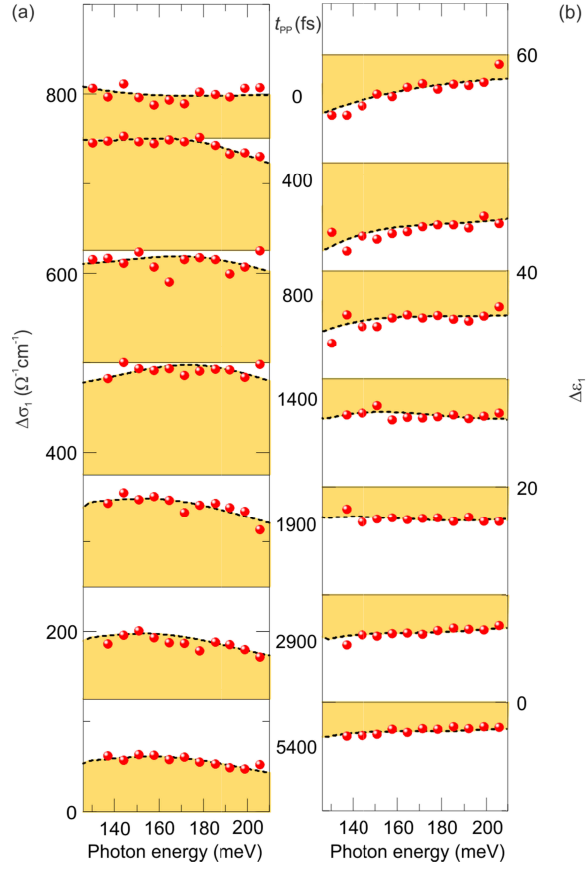

**Figure 3.** Pump-induced changes of the real parts of the optical conductivity $\Delta\sigma_1$ (a) and the dielectric function $\Delta\varepsilon_1$ (b) of the photoexcited WSe$_2$ monolayer as a function of the photon energy, for several pump delays $t_{PP}$ after non-resonant excitation. The pump fluence is set to 38 μJ/cm². Red spheres denote the experimental data and the black dashed curves represent the results by the Drude-Lorentz model fitting simultaneously $\Delta\sigma_1$ and $\Delta\varepsilon_1$.

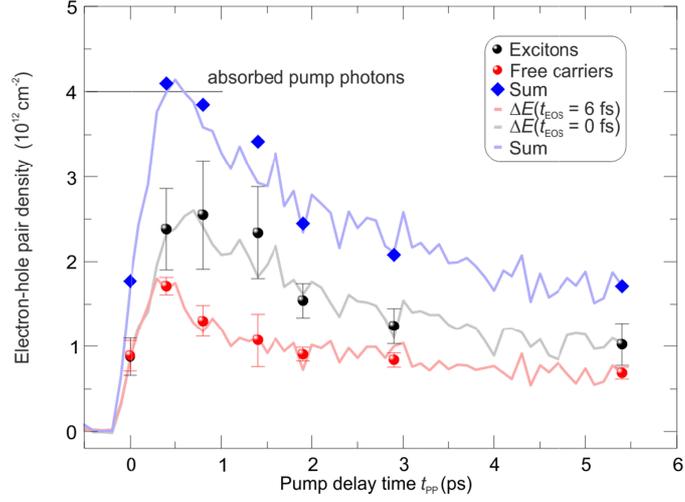

**Figure 4.** Absolute electron-hole pair densities extracted from the data shown in Figure 3, using the Drude-Lorentz model, including 1s excitons $n_X$ (black spheres), unbound electron-hole-pairs $n_{FC}$ (red spheres), and the total density of the two contributions (blue diamonds) as a function of the delay time $t_{PP}$ after non-resonant excitation. The error bars represent the 95% confidence intervals of the fitting parameters. In addition, pump-induced changes $\Delta E$ recorded at fixed electro-optic sampling times of $t_{EOS} = 0$ fs (gray solid line) and 6 fs (red solid line) as a function of $t_{PP}$ are presented. These changes are proportional to the exciton and plasma densities, respectively, as discussed in the text. The horizontal solid line marks the number of absorbed pump photons as estimated from the applied pump fluence and the optical absorbance.

1. Splendiani, A; Sun, L.; Zhang, Y.; Li, T.; Kim, J.; Chim, C.-Y.; Galli, G.; Wang, F. *Nano Lett.* **2010**, 10, 1271-1275.

2. Mak, K. F.; Lee, C.; Hone, J.; Shan, J.; Heinz, T. F. *Phys. Rev. Lett.* **2010**, 105, 136805.

3. Tonndorf, P.; Schmidt, R.; Böttger, P.; Zhang, X.; Börner, J.; Liebig, A.; Albrecht, M.; Kloc, C.; Gordan, O.; Zahn, D. R. T.; de Vasconcellos, S. M.; Bratschitsch, R. *Opt. Express* **2013**, 21, 4908-4916.

4. Moody, G.; Dass, C. K.; Hao, K.; Chen, C.-H.; Li, L.-J.; Singh, A.; Tran, K.; Clark, G.; Xu, X.; Berghäuser, G.; Malic, E.; Knorr, A.; Li. X. *Nat. Commun.* **2015**, 6, 8315.

5. Xu, X.; Yao, W.; Xiao, D.; Heinz, T. F. *Nat. Phys.* **2014**, 10, 343-350.

6. Wang, G.; Glazov, M. M.; Robert, C.; Amand, T.; Marie, X.; Urbaszek, B. *Phys. Rev. Lett.* **2015**, 115, 117401.

7. Plechinger, G.; Nagler, P.; Arora, A.; Schmidt, R.; Chernikov, A.; del Águila, A. G.; Christianen, P. C.M.; Bratschitsch, R.; Schüller, C.; Korn, T. *Nat. Commun.* **2016**, 7, 12715.

8. Wang, Q. H.; Kalantar-Zadeh, K.; Kis, A.; Coleman, J. N.; Strano, M. S. *Nat. Nanotechnol.* **2012**, 7, 699-712.

9. Britnell, L.; Ribeiro, R. M.; Eckmann, A.; Jalil, R.; Belle, B. D.; Mishchenko, A.; Kim, Y.-J.; Gorbachev, R. V.; Georgiou, T.; Morozov, S. V.; Grigorenko, A. N.; Geim, A. K.; Casiraghi, C.; Castro Neto, A. H.; Novoselov, K. S. *Science* **2013**, 340, 1311-1314.

10. Jariwala, D.; Sangwan, V. K.; Lauhon, L. J.; Marks, T. J.; Hersam, M. C. *ACS Nano* **2014**, 8, 1102-1120.

11. Koppens, F. H. L.; Mueller, T.; Avouris, Ph.; Ferrari, A. C.; Vitiello, M. S.; Polini, M. *Nat. Nanotechnol.* **2014**, 9, 180-793.

12. Berkelbach, T. C.; Hybertsen, M. S.; Reichman, D. R. *Phys. Rev. B* **2013**, 88, 045318.

13. Qiu, D. Y.; da Jornada, F. H.; Louie, S. G. *Phys. Rev. Lett.* **2013**, 111, 216805.

14. Zhang, C.; Johnson, A.; Hsu, C.; Li, L.; Shih, C. *Nano Lett.* **2014**, 14, 2443-2447.

15. Ye, Z.; Cao, T.; O'Brien, K.; Zhu, H.; Yin, X.; Wang, Y.; Louie, S. G.; Zhang, X. *Nature* **2014**, 513, 214-218.

16. He, K.; Kumar, N.; Zhao, L.; Wang, Z.; Mak, K. F.; Zhao, H.; Shan, J. *Phys. Rev. Lett.* **2014**, 113, 026803.

17. Chernikov, A.; Berkelbach, T. C.; Hill, H. M.; Rigosi, A.; Li, Y.; Aslan, O. B.; Reichman, D. R.; Hybertsen, M. S.; Heinz, T. F. *Phys. Rev. Lett.* **2014**, 113, 076802.

18. Ugeda, M. M.; Bradley, A. J.; Shi, S.-F.; da Jornada, F. H.; Zhang, Y.; Qiu, D. Y.; Ruan, W.; Mo, S.-K.; Hussain, Z.; Shen, Z.-X.; Wang, F.; Louie, S. G.; Crommie, M. F. *Nat. Mater.* **2014**, 13, 1091-1095.

19. Poellmann, C.; Steinleitner, P.; Leierseder, U.; Nagler, P.; Plechinger, G.; Porer, M.; Bratschitsch, R.; Schüller, C.; Korn, T.; Huber, R. *Nat. Mater.* **2015**, 14, 889-893.

20. Zhu, C. R.; Zang, K.; Glazov, M.; Urbaszek, B.; Amand, T.; Ji, Z. W.; Liu, B. L.; Marie, X. *Phys. Rev. B* **2014**, 90, 161302(R).

21. Chernikov, A.; Ruppert, C.; Hill, H. M.; Rigosi, A. F.; Heinz, T. F. *Nat. Photon.* **2015**, 9, 466-470.

22. Kumar, N.; Cui, Q.; Ceballos, F.; He, D.; Wang, Y.; Zhao, H. *Phys. Rev. B* **2014**, 89, 125427.

23. Mak, K. F.; He, K.; Lee, C.; Leem, G. H.; Hone, J.; Heinz, T. F.; Shan, J. *Nat. Mater.* **2013**, 12, 207-211.

24. You, Y.; Zhang, X.-X.; Berkelbach, T. C.; Hybertsen, M. S.; Reichman, D. R.; Heinz, T. F. *Nat. Phys.* **2015**, 11, 477-481.

25. Lagarde, D.; Bouet, L.; Marie, X.; Zhu, C. R.; Liu, B. L.; Amand, T.; Tan, P.H.; Urbaszek, B. *Phys. Rev. Lett.* **2014**, 112, 047401.

26. Wang, H.; Zhang, C.; Chan, W.; Manolatou, C.; Tiwari, S.; Rana, F. *Phys. Rev. B* **2016**, 93, 045407.

27. Zimmermann, J. E.; Mette, G.; Höfer, U. *arXiv* **2016**, 1608.03434v2.

28. Robert, C.; Lagarde, D.; Cadiz, F.; Wang, G.; Lassagne, B.; Amand, T.; Balocchi, A.; Renucci, P.; Tongay, S.; Urbaszek, B.; Marie, X. *Phys. Rev. B* **2016**, 93, 205423.

29. Selig, M.; Berghäuser, G.; Raja, A.; Nagler, P.; Schüller, C.; Heinz, T. F.; Korn, T.; Chernikov, A.; Malic, E.; Knorr, A. *arXiv* **(2016)**, 1605.03359v1.

30. Ceballos, F.; Cui, Q.; Bellus, M. Z.; Zhao, H. *Nanoscale* **2016**, 8, 11681-11688.

31. Huber, R.; Tauser, F.; Brodschelm, A.; Bichler, M.; Abstreiter, G.; Leitenstorfer, A. *Nature* **2001**, 414, 286-289.

32. Kaindl, R. A.; Carnahan, M. A.; Hägele, D.; Lövenich, R.; Chemla, D. S. *Nature* **2003**, 423, 734-738.

33. Leinß, S.; Kampfrath, T.; v. Volkmann, K.; Wolf, M.; Steiner, J. T.; Kira, M.; Koch, S. W.; Leitensdorfer, A.; Huber, R. *Phys. Rev. Lett.* **2008**, 101, 246401.

34. Porer, M.; Leierseder, U.; Ménard, J.-M.; Dachraoui, H.; Mouchliadis, L.; Perakis, I. E.; Heinzmann, U.; Demsar, J.; Rossnagel, K.; Huber, R. *Nat. Mater.* **2014**, 13, 857-861.

35. Ménard, J.-M.; Poellmann, C.; Porer, M.; Leierseder, U.; Galopin, E.; Lemaitre, A.; Amo, A.; Bloch, J.; Huber, R. *Nat. Commun.* **2014**, 5, 4648.

36. Cha, S.; Sung, J. H.; Sim, S.; Park, J.; Heo, H.; Jo, M.-H.; Choi, H. *Nat. Commun.* **2016**, 7, 10768.

37. Kubouchi, M.; Yoshioka, K.; Shimano, R.; Mysyrowicz, A.; Kuwata-Gonokami, M. *Phys. Rev. Lett.* **2005**, 94, 016403.

38. Huber, R.; Schimd, B. A.; Ron Shen, Y.; Chemla, D. S.; Kaindl, R. A. *Phys. Rev Lett.* **2006**, 96, 017402.

39. Castellanos-Gomez; A.; Buscema, M.; Molenaar, R.; Singh, V.; Janssen, L.; van der Zant, H. S. J.; Steele, G. A. *2D Mater.* **2014**, 1, 011002.

40. Li, Y; Chernikov, A.; Zhang, X.; Rigosi, A.; Hill, H. M.; van der Zande, A. M.; Chenet, D. A.; Shih, E.-M.; Hone, J.; Heinz, T. F. *Phys. Rev. B* **2014**, 90, 205422.

41. Kormányos, A.; Burkard, G.; Gmitra, M.; Fabian, J.; Zólyomi, V.; Drummond, N. D.; Fal'ko, V. *2D Mater.* **2015**, 2, 022001.

42. Kaindl, R. A.; Hängele, D.; Carnahan, M. A.; Chemla, D. S. *Phys. Rev. B* **2009**, 79, 045320.

43. Singh, A; Moody, G.; Tran, K.; Scott, M. E.; Overbeck, V.; Berghäuser, G.; Schaibley, J.; Seifert, E. J.; Pleskot, D. N.; Gabor, M.; Yan, J.; Mandrus, D. G.; Richter, M.; Malic, E.; Xu, X.; Li. X. *Phys. Rev. B* **2016**, 93, 041401.

44. Zhang , X.-X.; You, Q.; Yang, S.; Zhao, F.; Heinz, T. F. *Phys. Rev. Lett.* **2015**, 115, 257403.

**Acknowledgements**

The authors thank Andor Kormányos, Christoph Pöllmann, and Michael Porer for helpful discussions and Martin Furthmeier for technical assistance. This work was supported by the European Research Council through ERC grant 305003 (QUANTUMsubCYCLE) and by the Deutsche Forschungsgemeinschaft (DFG) through Research Training Group GK1570 and project grant KO3612/1-1. AC gratefully acknowledges funding from the Deutsche Forschungsgemeinschaft through the Emmy Noether Programme (CH1672/1-1).

**Supporting Information**

- (i) Optical pump and mid-infrared probe pulses
- (ii) Field-resolved optical pump/mid-infrared probe spectroscopy
- (iii) Degradation of the $WSe_2$ monolayer
- (iv) Temporal evolution of the field-resolved time-domain data
- (v) Temporal evolution of the fitting parameters